\documentclass[lettersize,journal]{IEEEtran}
\pdfoutput=1
\usepackage{amsmath,amsfonts}
\usepackage{amssymb}
\usepackage{mathtools}
\usepackage{amsthm}
\usepackage{algorithmic}
\usepackage{algorithm}
\usepackage{array}
\usepackage{booktabs}
\usepackage{float}
\usepackage{textcomp}
\usepackage{stfloats}
\usepackage{url}
\usepackage{verbatim}
\usepackage{graphicx}
\usepackage{subcaption}
\usepackage{cite}
\usepackage{nicefrac}       % compact symbols for 1/2, etc.
\usepackage{tikz}
\usetikzlibrary{shapes.geometric}
\usepackage[showseconds=false,showzone=false]{datetime2}

\usepackage{mathtools}
\DeclarePairedDelimiterX{\norm}[1]{\lVert}{\rVert}{#1}

\usepackage{array}
\usepackage{multirow}

\def\a{{\mathbf a}}

\def\x{{\mathbf x}}
\def\f{{\mathbf f}}
\def\g{{\mathbf g}}
\def\r{{\mathbf r}}

\def\w{{\mathbf w}}
\def\L{{\cal L}}

\allowdisplaybreaks
\begin{document}

\title{A Theoretical Framework for \\Acoustic Neighbor Embeddings}

\author{Woojay Jeon% <-this % stops a space
\thanks{Woojay Jeon is with Apple in Cupertino, California, U.S.A. (e-mail: woojay@apple.com)}
}% <-this % stops a space

% % \thanks{This paper was produced by the X. They are in Y.}% <-this % stops a space
% \thanks{Manuscript received xxxx; revised xxxx.}}

% \markboth{Submission to XXXX,~Vol.~X, No.~Y, WWWW~ZZZZ}%
% {Shell \MakeLowercase{\textit{et al.}}: Submission Pre-Print}

\markboth{Manuscript Preprint}%
{Shell \MakeLowercase{\textit{et al.}}: Submission Pre-Print}


\maketitle

\begin{abstract}
This paper provides a theoretical framework for interpreting acoustic neighbor embeddings, which are representations of the phonetic content of variable-width audio or text in a fixed-dimensional embedding space.
A probabilistic interpretation of the distances between embeddings is proposed, based on a general quantitative definition of phonetic similarity between words.
This provides us a framework for understanding and applying the embeddings in a principled manner.
Theoretical and empirical evidence to support an approximation of uniform cluster-wise isotropy are shown, which allows us to reduce the distances to simple Euclidean distances.
Four experiments that validate the framework and demonstrate how it can be applied to diverse problems are described.
Nearest-neighbor search between audio and text embeddings can give isolated word classification accuracy that is identical to that of finite state transducers (FSTs) for vocabularies as large as 500k.
Embedding distances give accuracy with 0.5\% point difference compared to phone edit distances in out-of-vocabulary word recovery, as well as producing clustering hierarchies identical to those derived from human listening experiments in English dialect clustering.
The theoretical framework also allows us to use the embeddings to predict the expected confusion of device wake-up words.
All source code and pretrained models are provided.
\end{abstract}

\begin{IEEEkeywords}
Acoustic neighbor embeddings, word embeddings
\end{IEEEkeywords}

\section{Introduction}\label{sec:introduction}

\IEEEPARstart{V}{ector} representations of words, called word embeddings, have been proven to be useful in diverse computational settings because any word of arbitrary length can be represented in a numeric form of constant length.
An \emph{acoustic} word embedding \cite{bengio14_interspeech,levin_2013,settle17_interspeech}, in particular, captures how a word sounds, as opposed to a semantic word embedding that captures what a word means.
The distance between any two embeddings represents their acoustic similarity, and has been applied to query-by-example search \cite{settle17_interspeech}, lattice rescoring in automatic speech recognition (ASR) \cite{bengio14_interspeech}, and keyword spotting \cite{7178970}.
A number of different embedding methods have been proposed in the literature.
These include linear and nonlinear embedding transformations \cite{levin_2013}, extraction of embeddings from a sequence-to-sequence speech recognizer \cite{palaskar-2019}, transformation via convolutional neural networks trained with contrastive loss \cite{7472619}, likewise with recurrent neural networks \cite{7846310}, and unsupervised methods using recurrent audioencoders \cite{8736337,9360516}.
Of particular interest to us is the training of an audio embedding network and a text embedding network in tandem such that an audio recording of a word would map to the same (or close) location in the embedding space as its corresponding text \cite{he-2017}.

Fundamental questions remain on how exactly acoustic word embeddings from neural networks, which specifically aim to represent speech, can model \emph{phonetic similarity} \cite{ladefoged-1969-measurement,VITZ1973373,kessler-2005,HAHN2005227}, and what the distances between embeddings truly represent.
While a comprehensive discussion of all the different meanings of phonetic similarity is beyond the scope of this paper, we are largely interested in two perspectives of phonetic similarity: 1. The \emph{acoustic} similarity in terms of what the speech signals are, and 2. The \emph{perceptual} similarity in terms of how the speech signals are perceived by humans.
In either case, it is often taken for granted that acoustic word embeddings model such similarity in some way, with little further attempt at theoretical interpretation and illumination that could enhance our understanding.
It is probably difficult to definitively show how an abstract space of vectors posited by a neural network coincides with a mathematical or psychoacoustic definition of phonetic similarity.
However, in this paper we will show that with some quantitative assumptions and approximations, a connection can be established that allows us to gain better understanding on the interpretation and usage of an acoustic word embedding, and apply it to various problems in a principled manner with greater clarity and confidence.

In particular, we will concern ourselves with a specific type of embedding learned via supervised training that we call \emph{acoustic neighbor embeddings} \footnote{Previously appeared in a preprint article in 2020 \cite{jeon-ane-2020}}.
The embedding is based on a modification of stochastic neighbor embedding \cite{hinton-beautiful}, and has been applied to query rewriting in dialogue systems \cite{nguyen21_interspeech}, embedding-matching automatic speech recognition (ASR) \cite{yen-2023}, and retrieval augmented correction of named entity errors \cite{pusateri2024retrievalaugmentedcorrectionnamed}.
Like \cite{he-2017}, our system gives rise to two broad types of embedders: An audio embedder $f(\cdot)$ that transforms audio, and a text embedder $g(\cdot)$ that transforms text.
In our system, we further divide the text embedder into a phone embedder that transforms a sequence of phones (which we also refer to as ``pronunciation''), and a grapheme embedder that transforms a sequence of graphemes. \footnote{We could furthermore have separate embedders for written-form graphemes (``I paid \$5'') and spoken-form graphemes (``I paid five dollars''), but here we are tied to whatever text is provided in Libriheavy, which can mostly be regarded as spoken-form. It should also be noted that throughout this paper, a ``word'' can be any arbitrary string, such as ``New York'' or ``A midsummer night's dream,'' because we regard the space character (`` '') as another grapheme.}

We will begin our discussion with a fundamental quantitative definition of phonetic similarity.
Next, we will describe how acoustic neighbor embeddings are trained, and show how they fit into our aforementioned definition.
We give theoretical and empirical evidence that supports an approximation of constant cluster-wise isotropy, which allows us to use a simplified equation that justifies nearest-neighbor search based on Euclidean distances.
A fundamental interpretation of the distance between an audio embedding and a text embedding, and the distance between two text embeddings, is proposed.
Finally, we describe four different experiments using the Libriheavy\cite{kang2023libriheavy} corpus in word classification, out-of-vocabulary(OOV) word recovery, dialect clustering, and wake-up word confusion prediction that demonstrate how the framework can be applied in diverse problem settings and also serve as a validation of the proposed framework.
All source code and pretrained models \footnote{\texttt{github.com/apple/ml-acn-embed}} are provided so that experiments can be replicated and models can be used for future research.

\section{A Fundamental Definition\\of Phonetic Similarity}

While the definition we chose for our framework is more easily understood as an acoustic similarity, we will later argue that it can also be a perceptual similarity.

\subsection{A hypothetical experiment}\label{sec:hypothetical}

\begin{figure}[t]
     \centering
     \begin{subfigure}{0.5\textwidth}
          \centering
          \includegraphics[width=0.6\textwidth]{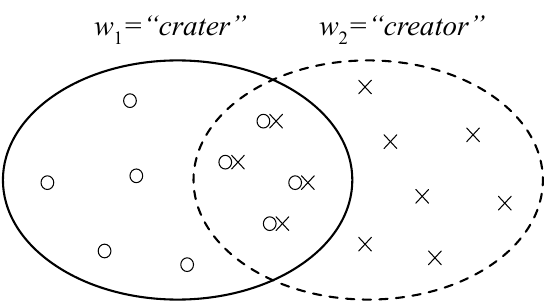}
          \caption{Example Venn diagram showing $n=9$ recordings of ``crater'' and $m=11$ recordings ``creator,'' out of which there are $p=4$ identical pairs. The phonetic similarity between ``crater'' and ``creator'' is therefore $4/20=0.2$.}
          \label{fig:venn}
     \end{subfigure}%
     \hfill
     \vspace{0.1in}
     \begin{subfigure}{0.5\textwidth}
          \centering
          \includegraphics[width=0.9\textwidth]{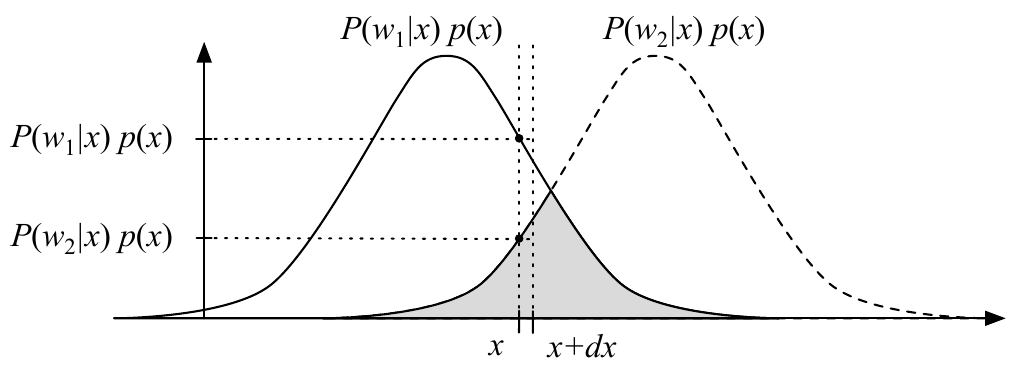}
          \caption{Abstraction of Fig. \ref{fig:venn} in probabilistic terms, where the audio recordings are projected onto an acoustic feature space $X$ (which is the set of all real scalars in this example). For audio recordings lying in $(x, x+dx)$, the proportion of identical pairs is $P(w_2|x)p(x)dx$.}
          \label{fig:overlap}
     \end{subfigure}
     \caption{A definition of phonetic similarity between two words via a hypothetical physical experiment that is abstracted in probabilistic terms}
     \label{fig:similarity}
\end{figure}

We consider the hypothetical experiment concerning the two words $w_1=$ ``crater'' and $w_2=$ ``creator''.
We visit every English speaker in the universe and ask them to choose and speak one of the two words.
The utterance is recorded, and labeled with the chosen word.
If enough recordings are collected from enough speakers, we will eventually encounter a recording of ``crater'' that is exactly identical to another recording of ``creator'' (e.g. assume the recordings are in WAV format and are compared bit-by-bit).
We can draw a Venn Diagram as in Fig. \ref{fig:venn}, where such identical pairs are placed in the overlapping area of the two ovals.
If we have $n$ recordings of ``crater'', $m$ recordings of ``creator'', and $p$ identical pairs, we define the phonetic similarity between ``crater'' and ``creator'' as the ratio between the number of identical pairs and the total number of recordings:\footnote{This ratio is like the Jaccard similarity, but using pairs instead of total counts in the numerator}
\begin{equation}\label{eq:def1}
\textrm{Phonetic Similarity} \triangleq \frac{p}{n+m}.
\end{equation}
Since $p$ cannot be negative, and is at most $(n+m)/2$, the similarity in \eqref{eq:def1} is limited to the range $[0, \frac{1}{2}]$.
When the two words are totally different, no identical pair will exist, so the similarity metric will be $0$.
When the two words are the same (e.g. both are ``crater'') and $n=m$, all recordings will form identical pairs in the limit, so the similarity will be the maximum value $\frac{1}{2}$.

Since such an experiment cannot be carried out practically, we abstract the experiment in probabilistic terms.
If we assume the audio recordings are projected onto a 1-dimensional acoustic feature space $X$ where each sound is represented by a point $x \in X$, we can consider the audio recordings in a small range $(x, x+dx)$ in Fig. \ref{fig:overlap}.
The number of recordings in this range is $(n+m)p(x)dx$. Out of this number, the proportion of recordings for ``crater'' is $P(w_1|x)$, so the number of recordings of ``crater'' in the given range is $(n+m)P(w_1|x)p(x)dx$.
Likewise, the number for ``creator'' is $(n+m)P(w_2|x)p(x)dx$. 
The number of identical pairs is the lower of the two numbers, which, in the case of the example location in Fig. \ref{fig:overlap}, is the latter.
Since we want to do this over all possible values of $x$ and compute the sum, it follows that the phonetic similarity defined in \eqref{eq:def1} is the area of overlap between $P(w_1|x)p(x)$ and $P(w_2|x)p(x)$ over $X$, shown by the gray shade in Fig. \ref{fig:overlap}.
Hence, we define the phonetic similarity $s(w_1, w_2; \lambda)$, where $\lambda$ is the set of all model parameters, as
\begin{multline}\label{eq:def2}
s(w_1, w_2; \lambda) \triangleq \\
\int_X \min \left\{ P(w_1|\x)p(\x), P(w_2|\x)p(\x) \right\} d\x,
\end{multline}
where we have now generalized $x$ to a multidimensional variable $\x$, and we note that $P(w|\x)p(\x)$ can also be written as $p(\x|w)P(w)$.

Eq. \eqref{eq:def2} is the well-known Bayes Error Rate \cite{duda-2000} for the two-class classification problem, where the two classes are $w_1$ and $w_2$. \footnote{The Bayes Error Rate is easily extensible to $n>2$ classes, but in this paper we will deal only with pairwise comparisons, and do multiple pairwise comparisons when there are $n>2$ classes (e.g. see Sec. \ref{sec:dialect_analysis}).}
The more similar the acoustic observations of $w_1$ and $w_2$, the harder it is to discriminate between them, and therefore the higher the classification error rate.
Two caveats are worth noting when discussing the Bayes Error Rate: 1. The features must be sufficiently discriminative (e.g. if $X$ is simply signal energy, there would always be very large overlap in Fig. \ref{fig:overlap} for any pair of words and the similarity metric would be useless), and 2. we almost never know the true probability distributions in Fig. \ref{fig:overlap}, and all the distributions we use in practice are only estimates.
Even if we are able to exactly evaluate the integral in \eqref{eq:def2}, we would be obtaining an \emph{estimated} similarity.

Note that \eqref{eq:def1} and \eqref{eq:def2} are affected by the choice of the total counts $n$ and $m$, which is equivalent to choice of priors $P(w_1)$ and $P(w_2)$.
The choice of priors is application-dependent.
If the speakers in our hypothetical experiment are allowed to freely choose between the two words, the priors are unigram stochastic language model (LM) probabilities.
In most cases, however, it is reasonable to assume equal priors, i.e., we flip a fair coin to decide which word should be spoken by each hypothetical speaker such that, in the limit, $n=m$ and $P(w_1)=P(w_2)=\frac{1}{2}$.
In all the applications discussed in the rest of this this paper, we will assume equal priors in \eqref{eq:def1} and \eqref{eq:def2}.
Note that LMs can still be later combined with our framework in a different manner (e.g. Section \ref{sec:wakeup}).

\subsection{Approximation of phonetic similarity}\label{sec:approximation-of-phonetic-similarity}
In general, the similarity measure in \eqref{eq:def2} is not analytically tractable (e.g. one could use kernel density estimation methods \cite{MCDERMOTT2004107}).
A well-known upper bound is the Bhatacharyya Bound \cite{duda-2000}, which we use as an approximation:
\begin{equation}\label{eq:bound}
s(w_1, w_2; \lambda) \le \sqrt{P(w_1)P(w_2)} \int_X \sqrt{p(\x|w_1)p(\x|w_2)}d\x.
\end{equation}
A closed form equation exists for the RHS of \eqref{eq:bound} when $p(\x|w_1)$ and $p(\x|w_2)$ are Gaussian with mean $\mathbf{m_1}$ and $\mathbf{m_2}$ and covariances $C_1$ and $C_2$, respectively \cite{duda-2000}:
\begin{equation}\label{eq:gauss_bound}
\resizebox{1\hsize}{!}{$\frac{1}{2} \exp \Big[ -\frac{1}{8} (\mathbf{m}_2-\mathbf{m}_1)^T \left\{ \frac{C_1 + C_2}{2} \right\} ^ {-1}  (\mathbf{m}_2-\mathbf{m}_1) 
-\frac{1}{2} \ln \frac{\left|\frac{C_1+C_2}{2}\right|} {\sqrt{|C_1||C_2|}}
\Big]$},
\end{equation}
where we have also assumed equal priors for $w_1$ and $w_2$.
Furthermore, if the two distributions are \emph{equally isotropic}, i.e., both covariance matrices equal $\sigma^2 I$ where $\sigma^2$ is a constant variance and $I$ is the identity matrix, the Bhatacharrya Bound simplifies to
\begin{equation}\label{eq:iso_bound}
\frac{1}{2} \exp \left[
-\frac{1}{8 \sigma^2} \big\|
\mathbf{m}_1 - \mathbf{m}_2
\big\|^2
\right].
\end{equation}

\subsection{Relation to asymmetric confusion probabilities}
We digress for a moment to note that, compared to the symmetric similarity in \eqref{eq:def2}, a popular notion of asymmetric phonetic confusion \cite{miller-1955} is usually formulated as a conditional probability \cite{burridge} where given an utterance labeled $w_2$, we ask what is the probability of the utterance being classified as $w_1$.
A phonetic confusion matrix can be constructed from estimates of the conditional probabilities, via physical human listening experiments \cite{miller-1955} or by gathering statistical data on an ASR's misrecognitions \cite{Srinivasan2000}.
However, such methods are usually only limited to pairs of single phones, and are hard to scale directly to arbitrary sequences of phones without relying on further heuristics (see Section \ref{sec:phonetic_match}).

In the case of the two words in Fig. \ref{fig:overlap}, it is straightforward to see that the left half of the area of overlap will be the probability of the utterance being classified as $w_1$ given that it is $w_2$, and the right half of the overlap will be the probability of classification as $w_2$ given that the utterance is $w_1$.
If we assume equal priors and variance, both halves are also equal, and the conditional confusion is the same as our similarity metric in \eqref{eq:def2}.
Note, however, there can be another subtly different way of measuring the confusion, which is to measure the probability that an audio signal labeled as $w_2$ is labeled as $w_1$ in the training data.
In this case, we have a different formulation:
\begin{align}
P(w_1|w_2)&=\int_X P(w_1|\x)p(\x|w_2)d\x. \label{eq:confusion}
\end{align}
Even with the assumption of Gaussianity, it is hard to find a closed form approximation for \eqref{eq:confusion}.
In this paper, we will not concern ourselves with asymmetric confusion probability and focus only on the similarity metric in \eqref{eq:bound}, \eqref{eq:gauss_bound}, and \eqref{eq:iso_bound}.

\section{Acoustic Neighbor Embeddings}

\subsection{Review of stochastic neighbor embeddings}

Stochastic neighbor embedding (SNE) \cite{hinton-beautiful} is a method of reducing the dimensions of vectors in a set of data while preserving the relative distances between vectors.
Given a set of $N$ coordinates $\{ \a _1, \cdots, \a _N\}$ where each $\a_i$ is a multidimensional vector, we train a function $f(\cdot)$ that maps each $\a _i$ to another vector of lower dimensions $\f _i$ where the relative distances among the $\a _i$'s are preserved among the corresponding $\f _i$'s.
The distance between two points $\a _i$ and $\a _j$ in the input space is defined as:
\begin{equation}\label{eq:euclidean_dist}
d^2_{ij} = \frac{ || \a _i  - \a _j || ^2 }{2\sigma^2_i},
\end{equation}
for some scale factor $\sigma_i$.
The probability of $\a _i$ ``choosing'' $\a _j$ as its neighbor in the input space is defined as:
\begin{equation}\label{eq:neighbor_prob}
p_{ij} = \frac{\exp \left( -d_{ij}^2 \right) }{\sum_{k \ne i} \exp \left( -d_{ik}^2 \right) }.
\end{equation}
In the embedding space, a corresponding ``induced'' probability is defined as
\begin{equation}\label{eq:induced_prob}
q_{ij} = \frac{\exp \left( - || \f_i - \f_j ||^2 \right) }{\sum_{k \ne i} \exp \left( - || \f_i - \f_k ||^2 \right) }.
\end{equation}
The loss function for training $f$ is the Kullback-Leibler divergence between the two distributions:
\begin{equation}\label{eq:kl_dist}
\L_f = \sum_{i, j} p_{ij} \log \frac{p_{ij}}{q_{ij}},
\end{equation}
which can be differentiated to obtain
\begin{equation}\label{eq:sne_gradient}
\frac{\partial \L_{f}}{\partial \f_i} = 2 \sum_j (\f_i - \f_j)(p_{ij}- q_{ij} + p_{ji} - q_{ji}).
\end{equation}

\subsection{Audio embedder training for acoustic neighbor embeddings}
\label{sec:audio_embedder_training}

\begin{figure}
     \centering
     \begin{subfigure}{0.48\textwidth}
          \centering
          \includegraphics[width=0.91\textwidth]{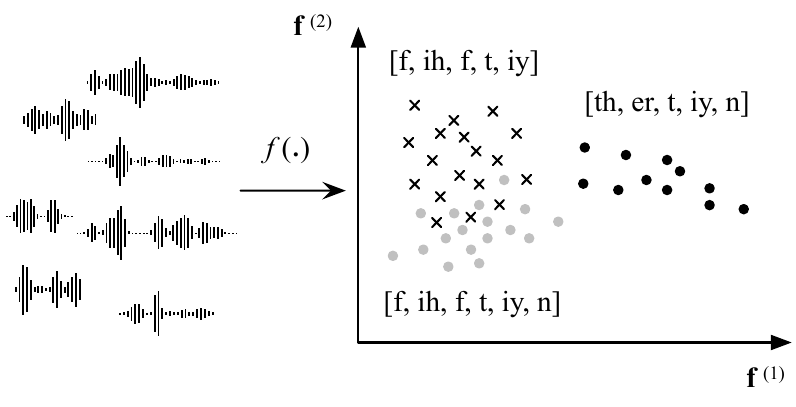}
          \caption{An audio embedder $f(\cdot)$ is trained to map different audio samples with the same label (phone sequence) to nearby locations. Since ``fifty'' and ``fifteen'' have similar sound, their clusters tend to overlap, whereas the embeddings for ``thirteen'' tend to be more detached from the rest.}
          \label{fig:f-training}
     \end{subfigure}%
     \hfill
     \vspace{0.1in}
     \begin{subfigure}{0.48\textwidth}
          \centering
          \includegraphics[width=0.91\textwidth]{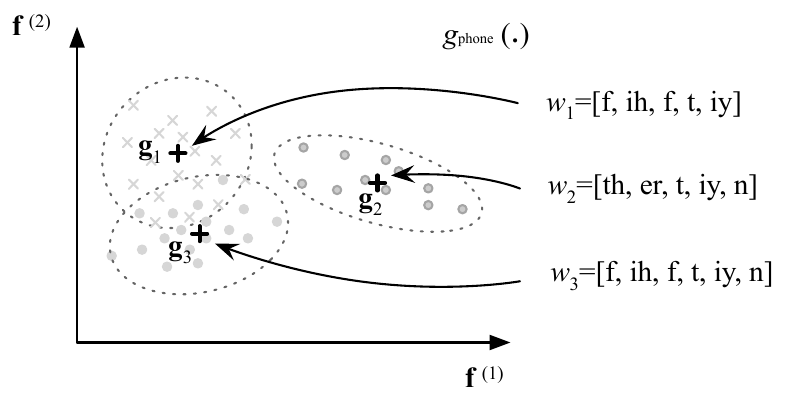}
          \caption{Once an $f(\cdot)$ embedder is trained, a text embedder $g_{\textrm {phone}}(\cdot)$ is trained for the phone sequences. Since a mean square error criterion is used, the text embeddings tend toward the cluster centroids.}
          \label{fig:g-training}
     \end{subfigure}
     \caption{Example 2-dimensional embedding space where an audio embedder and text embedder are trained over diverse audio samples for the words ``fifty'', ``fifteen'', and ``thirteen.''}
     \label{fig:training}
\end{figure}

Acoustic neighbor embeddings are obtained by training audio embedders using a loss function similar to that of stochastic neighbor embeddings, with the following two key differences:
\begin{enumerate}
     \item The transformation is applied to variable length sequences instead of vectors, such that instead of projecting a vector $\a$ to a vector $\f$ as in SNE, we project a \emph{sequence} of vectors $A=\{\a_1, \cdots, \a_T\}$ to a single vector $\f$.
     \item SNE has no notion of labels, only distances in the input space. We, on the other hand, have only labels, and assign binary distances in the input space based only on whether samples have the same label or not.
\end{enumerate}

Assuming a training data set of $N$ utterances of variable length, the $i$'th utterance is described by $(A_i, B_i, C_i)$ where $A_i$ is a sequence of acoustic feature vectors $[\a_1, \cdots, \a_T]$, $B_i$ is a sequence of phones (e.g. {\fontfamily{cmr}\selectfont[g, uh1, d, jh, aa1, b]}), and $C_i$ is a sequence of graphemes (e.g. {\fontfamily{cmr}\selectfont[g, o, o, d, \textunderscore, j, o, b]} where \textunderscore \hspace{0pt} indicates a space character).
$B_i$ and $C_i$ represent the reference transcription of $A_i$, and can all have different lengths.
Typically, only $C_i$ is originally available, and we obtain $B_i$ by force-aligning the audio against $C_i$ using a DNN-HMM hybrid ASR with a pronunciation dictionary.
The vectors in $A_i$ may be any known acoustic features such as MFCCs\cite{davis-1980} or filterbank outputs.
In our implementation, we used the outputs of a monophone DNN-HMM hybrid acoustic model -- which are the phone-wise posteriors at every frame, also called posteriorgram \cite{hazen-2009} -- as our acoustic features.
This allows the actual acoustic embedder network $f(\cdot)$ to be small, since speaker and channel normalization is already done by the DNN-HMM acoustic model.

Since each $A_i$ is a sequence, the Euclidean distance in \eqref{eq:euclidean_dist} is not applicable in our input space.
Instead, for each pair of points $A_i$ and $A_j$, we assign a binary distance based on whether their phone transcriptions $B_i$ and $B_j$ are exactly the same or not (see Section \ref{sec:binary} for further discussion on the use of a binary distance instead of the edit distance):
\begin{equation}\label{eq:dist}
d_{ij}=
\begin{cases}
0 &\text{if \(B_i = B_j\)} \\
\infty &\text{else}
\end{cases}.
\end{equation}
The distance in \eqref{eq:dist} results in the following input probability for \eqref{eq:neighbor_prob}:
\begin{equation}\label{eq:neighbor_prob2}
p_{ij}=
\begin{cases}
1/n_i &\text{if \(B_i = B_j\)} \\
0 &\text{else}
\end{cases},
\end{equation}
where $n_i$ is the number of utterances (other than the $i$'th) that have the same subword sequence $B_i$.
In the embedding space, we use the same induced probability $q_{ij}$ in \eqref{eq:induced_prob}, and we apply the same loss function $\L _f$ in \eqref{eq:kl_dist} to train the neural network.

Since our training data consists of millions of utterances, it is not possible to optimize the loss in \eqref{eq:kl_dist} over the entire data.
Instead, we divide the data into random, fixed-size \emph{microbatches}, where each microbatch of size $M$ has a random ``pivot'' sample $(A_0, B_0, C_0)$, and for the remaining $M-1$ samples there is at least one sample that has the same transcription as the pivot.
For further simplicity, instead of computing the loss in \eqref{eq:kl_dist} over all possible $i, j$ pairs in the microbatch, we fix $i$ to $0$, so that we only consider pairs that include the pivot.
For each microbatch, we have:
\begin{equation}\label{eq:kl_dist_simple}
\L_f = \sum_{j=1}^{M-1} p_{0j} \log \frac{p_{0j}}{q_{0j}},
\end{equation}
and the overall loss that we minimize (via minibatch training) is the average microbatch loss.
We show examples of this construction in Appendix \ref{sec:app-training}.

\subsection{Text embedder training}

Once we have fully trained $f(\cdot)$, we train a text encoder(s) $g(\cdot)$ such that its output $\g_i$ for every subword sequence will match the output of $f(\cdot)$ for audio samples that have that subword sequence.
For a phone encoder, said subword sequence is $B_i$, whereas for a grapheme encoder, it is $C_i$.
We apply a mean square error loss for this purpose:
\begin{align}\label{eq:mse}
\begin{dcases}
\L_{g,\textrm{phone}} = \sum_{n=1}^N || g(B_n) - f(X_n) || ^2 \\
\L_{g,\textrm{grapheme}} = \sum_{n=1}^N || g(C_n) - f(X_n) || ^2.
\end{dcases}
\end{align}
In the ideal case, the mean square error criterion will cause the embedding for a given text to converge to the mean of all the audio embeddings with that text label in the training data:
\begin{align}\label{eq:mean}
\begin{dcases}
g_{\textrm{phone}}(B) =E[f(X)|B]=E[\f|B] \\
g_{\textrm{grapheme}}(C) =E[f(X)|C]=E[\f|C].
\end{dcases}
\end{align}
where $\f$ is the audio embedding trained in Section \ref{sec:audio_embedder_training}.

\section{Interpretation of the distances}

\subsection{Approximation of constant cluster-wise isotropy}

\begin{figure}[t]
    \centering
    \includegraphics[width=0.45\textwidth]{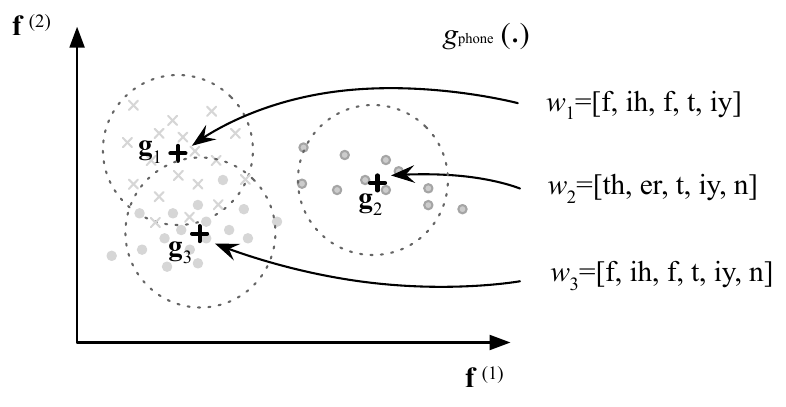}
    \caption{Fig. \ref{fig:g-training} where we now approximate each cluster to have a Gaussian distribution with equal isotropic covariances represented by the dotted circles.}
    \label{fig:g-isotropic}
\end{figure}

If we assume that each cluster of audio embeddings for a given word takes on a Gaussian distribution, we can directly apply \eqref{eq:bound} to compute the approximate phonetic similarity between words, where the distribution means are the text embeddings for each cluster.
However, it is generally hard to reliably estimate full covariances.
If we can make a further approximation that all the clusters are \emph{equally isotropic}, we can use the much simpler form in \eqref{eq:iso_bound}.
We are highly motivated to make this approximation because the reduction to a Euclidean distance simplifies computation and allows us to exploit many large-scale nearest neighbor search techniques \cite{locality-sensitive-hashing,garcia-2008,sun-2024}.
While it is difficult to prove formally how isotropic our embedding clusters are, in Appendix \ref{sec:app-isotropy} we show theoretical and experimental evidence that partially supports our approximation of equal cluster-wise isotropy \footnote{Not to be confused with \emph{global} isotropy \cite{mickus-etal-2024-isotropy}}.
We redraw Fig. \ref{fig:g-training} as Fig. \ref{fig:g-isotropic}, where each cluster of audio embeddings for each pronunciation has an isotropic distribution, all of equal size.

Strictly speaking, because the audio embedder is trained using the phone-based label in \eqref{eq:dist}, the isotropy assumption in \eqref{eq:iso_bound} only holds for the phone embedder, not the grapheme embedder.
To better ensure isotropy in the grapheme embeddings, one would need to train a new audio embedder using $C_i$ and $C_j$ in \eqref{eq:dist}.
However, every time a new audio embedder is trained, the embeddings can change completely.
Using the same audio embedder to train both the phone and grapheme embedders has the added benefit of making all three embedders consistent with each other, allowing applications to use phones and graphemes interchangeably.

\subsection{Two fundamental distances}

In summary, we characterize the Euclidean distance between embeddings as follows, where $\lambda$ represents the parameters for all our neural network embedders:
\begin{enumerate}
    \item When computing the $\L_2$ distance between an audio embedding $\f$ and a text embedding $\g$ for a word $w$, we are in effect evaluating a Gaussian acoustic likelihood:
    \begin{equation} \label{eq:fg_dist}
        p(\f | w; \lambda) = 
        \frac{1}{(\sqrt{2 \pi} \sigma)^d}
        \exp \left(
        -\frac{1}{2 \sigma^2}
        {\Big \|}\f - \g{\Big \|}^2
        \right).
    \end{equation}

    \item When computing the $\L_2$ distance between two text embeddings $\g_i$ and $\g_j$ for two words $w_i$ and $w_j$, respectively, we are in effect computing the phonetic similarity between $w_i$ and $w_j$:
    \begin{equation} \label{eq:gg_dist}
        s(w_i, w_j; \lambda) =
        \frac{1}{2}
        \exp \left(
        -\frac{1}{8 \sigma^2}
        {\Big \|}\g_i - \g_j{\Big \|}^2
        \right).
    \end{equation}

\end{enumerate}

In many applications, we do not need to know $\sigma$ and can drop most of the operations in \eqref{eq:fg_dist} and \eqref{eq:gg_dist}, which leaves us with only the $\mathcal{L}_2$ distances.

\subsection{Binary distances in the input space} \label{sec:binary}

Returning to the hypothetical experiment in Sec. \ref{sec:hypothetical} and the fundamental definition in \eqref{eq:def1}, we see there is no need to assign explicit distances and similarities in the input space in \eqref{eq:dist} and \eqref{eq:neighbor_prob2}.
The hypothetical experiment imposes only a binary label to each recording (e.g. each speaker simply declares their utterance as either ``crater'' or ``creator'', never ``75\% crater and 25\% creator''), and cares only about the proportion of times the same audio is labeled as ``crater'' versus ``creator.''
Likewise, in \eqref{eq:neighbor_prob2} the same audio input will sometimes be assigned a positive score for some word $w_i$ and sometimes a positive score for some word $w_j$.
As long as we have a sufficient amount of data, many such training samples will exist for the same audio input, and the samples' binary labels will ``compete'' against each other in \eqref{eq:gradient_plus}.
We will eventually reach some equilibrium state where the location of the embedding reflects the proportion of times the embedding belongs to each cluster.
In fact, we observed significantly worse experimental results when we tried to enforce some prenotion of similarity by setting $d_{ij}$ using dynamic time warping (DTW) between $X_i$ and $X_j$ (in a manner similar to \cite{hazen-2009}) or the edit distance between $B_i$ and $B_j$.
The simple binary distance in \eqref{eq:dist} always gave the best result.
A related observation was made in \cite{7472619} where a neural network trained on binary labels indicating whether a pair of words are the same or different outperformed dimension reduction techniques applied to DTW-based reference scores \cite{levin_2013}.

\subsection{Acoustic vs. perceptual similarity}

If the feature space $X$ in Eq. \eqref{eq:def2} is based on a model of perception (such as mel-frequency warping \cite{davis-1980} in the energy filterbanks in our acoustic frontend), it can be argued that the acoustic similarity measure is also a perceptual similarity.
Furthermore, it can be generally argued that any statistical speech model trained on human-transcribed or human-read speech is perceptual -- albeit to varying degrees -- since the training audio is already labeled and organized according to how it is perceived by humans.
Even if we used pure WAV bits to compute the similarity in \eqref{eq:def1}, the similarity function would have perceptual properties because the data samples and labels in Fig. \ref{fig:venn} come from humans.
If the speakers in the hypothetical experiment in Sec. \ref{sec:hypothetical} were aliens who perceived sound waves differently from humans, for instance, and as a consequence voiced the words differently, the similarity measure would result in a different value.
One experiment that supports our argument that our similarity is perceptual is described in Sec. \ref{sec:dialect_analysis}.

\section{Experiments}\label{sec:experiments}

\begin{table}[t]
     \caption{Embedder type used in each experiment, and summary of each experiment's purpose}
     \label{tab:embedder_types}
     \centering
     \begin{tabular}{clll}
     \toprule
     No.  & Experiment & Type(s)   & Purpose \\ 
     \midrule
     1    & Word            & Audio         & Validate \eqref{eq:fg_dist} and compare with \\
          & classification  & Phone         & conventional isolated-word ASR \\
          &                 & Grapheme      & \\
     \midrule
     2    & OOV word        & Phone         & Validate \eqref{eq:gg_dist} and compare with \\
          & recovery        &               & traditional edit-distance \\
     \midrule
     3    & Dialect         & Audio         & Validate \eqref{eq:mean} \& \eqref{eq:gg_dist}. Test a\\
          & clustering      &               & connection between embeddings \\
          &                 &               & and perceptual similarity\\
     \midrule
     4    & Wake-up word    & Grapheme      & Show another application of \eqref{eq:gg_dist}  \\
          & confusion       &               & and theoretical framework in \ref{sec:approximation-of-phonetic-similarity} \\
     \bottomrule
     \end{tabular}
\end{table}

In this section, we discuss four experiments using acoustic neighbor embeddings to demonstrate how to apply the proposed theoretical framework as well as provide evidence of its validity, as summarized in Tab. \ref{tab:embedder_types}.
Implementation details that are too numerous to describe here can be found in the source code.
We used the Libriheavy \cite{kang2023libriheavy} corpus for all audio data and corresponding transcriptions.
Where applicable, we also used the Kaldi \cite{Povey_ASRU2011} speech recognition toolkit, PyTorch \cite{10.5555/3454287.3455008}, the Carnegie Mellon Pronouncing Dictionary \cite{cmudict}, and the LibriSpeech \cite{7178964} pronunciation lexicon, grapheme-to-phone (G2P) model, and LMs \cite{openslr}.

\subsection{Data preparation and embedder training} \label{sec:training}

\begin{table}[h]
     \caption{Non-overlapping datasets extracted from Libriheavy ``large'' data}
     \label{tab:datasets}
     \centering
     \begin{tabular}{ccrl}
     \toprule
     Dataset & Hours(k)   & Utterances   & Purpose \\ 
     \midrule
     TR-A    & 10  & 2,632,239    & Train DNN-HMM acoustic models \\
     TR-B    &  1  & 262,944      & Train word embedders \\
     CV      &  1  & 263,122      & Cross-validation for stopping training \\
     EXP     &  1  & 263,431      & Word classification experiment \\
     \bottomrule
     \end{tabular}
\end{table}

From the Libriheavy ``large'' data, we randomly extracted four non-overlapping sets of utterances: \emph{TR-A}, \emph{TR-B}, \emph{CV}, and \emph{EXP}, as shown in Table. \ref{tab:datasets}.
Due to limitations in the G2P, we excluded any utterances containing Arabic or Roman numerals.
First, a conformer \cite{gulati-2020} DNN-HMM hybrid acoustic model \cite{bourlard-1994} that converts 80 mel filterbank features to 5,744 HMM-state-level framewise posterior scores was trained using \emph{TR-A} data and a cross-entropy loss.
The network had 10 heads, 23 layers, 300 internal embedding dimensions, a kernel size of 31, and a total 50M trainable parameters.
When used with the LibriSpeech \emph{3-gram.pruned.1e-7} \cite{openslr} LM without further rescoring, the hybrid AM had word error rates (WERs) of 4.9\% and 10.3\% on the Libriheavy ``test-clean'' and ``test-other'' data.\footnote{Earlier versions of this hybrid model were also used to scrub the ``large'' data so that utterances that failed force alignment against the transcriptions (usually due to mismatch between audio and text) would be discarded}
This hybrid AM was used to force-align the \emph{TR-A} training data to the reference transcriptions to obtain frame-wise monophone sequences, which were used to train a second conformer DNN-HMM hybrid acoustic model -- with identical configuration as the first model except the final linear layer -- that outputted 70 monophone posterior scores (69 non-silence ARPAbet phones \cite{openslr} + silence phone) and had a total 48M trainable parameters.
When used in the same manner as the first hybrid AM, the monophone AM gave WERs of 8.6\% and 14.8\% on ``test-clean'' and ``test-other,'' respectively (there was significant degradation due to the reduced outputs).
This monophone model served as the acoustic frontend for the proposed audio embedder.

\begin{figure}[t]
    \centering
    \includegraphics[width=0.40\textwidth]{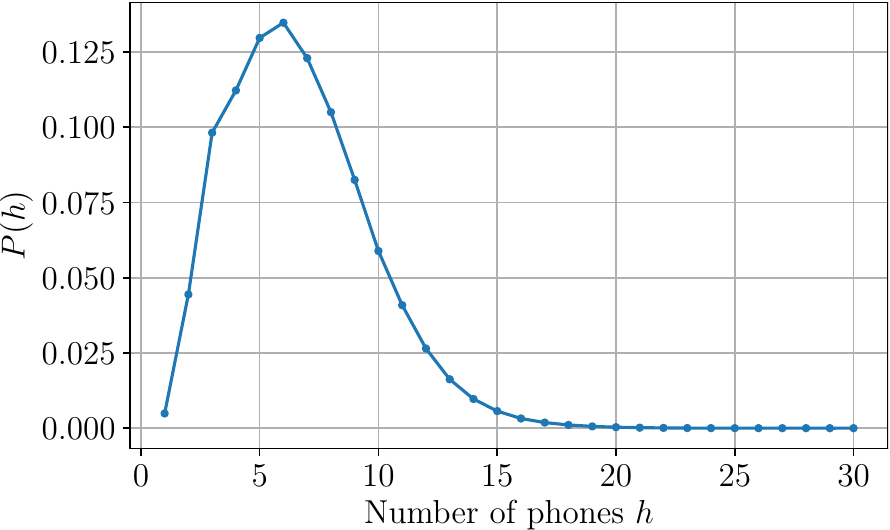}
    \caption{Normalized distribution of word pronunciation lengths (i.e., number of phones) used for encoder training.}
    \label{fig:pronlendist}
\end{figure}

To prepare training samples for the embedders, we estimated a distribution of input lengths for the English language using the LibriSpeech \emph{3-gram.pruned.1e-7} LM and G2P \cite{openslr}.
The probability of an utterance having a pronunciation of length $h$ is
\begin{equation}
P(h) = \sum_{\r \in R_h} \sum_{\w \in W_r} P(\r | \w) P(\w),
\end{equation}
where $\w$ is a sequence of words, $\r$ is a pronunciation, $W_r$ is the set of all words with pronunciation $\r$, and $R_h$ is the set of all pronunciations with length (number of phones) $h$. 
For simplicity, we assumed a uniform pronunciation probability $P(\r|\w)=1 / n_{\w}$ where $n_{\w}$ is the number of possible pronunciations for $\w$.
$P(\w)$ was obtained from the LM, limited only to word sequences for which explicit $n$-gram probabilities were available, and $W_r$ and $R_h$ were obtained from the pronunciation lexicon and G2P.
The resulting distribution of pronunciation lengths is in Fig. \ref{fig:pronlendist}.

The audio encoder was a bidirectional LSTM with two layers and 100 states per layer trained per Section \ref{sec:audio_embedder_training}.
An additional linear layer was applied to the last output of every sequence to produce the desired embedding dimensions.
\emph{Adam} \cite{kingma-2017} optimization was used with an initial learning rate of 0.001.
\emph{TR-B} was used as training data and \emph{CV} was used to stop training (the aforementioned hybrid acoustic models were used to obtain both datasets' word segmentations and monophone posteriors).
Every microbatch was randomly generated on-the-fly during training, with the segments chosen such that the distribution of lengths approximate that in Fig. \ref{fig:pronlendist}.
This was achieved by a greedy algorithm where a histogram of pronunciation lengths used so far during training was maintained.
Every time a training sample was needed, we compared the histogram with the desired distribution to find the most sample-deficient histogram bin, then looked for a random segment with the corresponding length.
Each microbatch had 160 utterances, and each minibatch had 32 microbatches.

\subsection{Experiment 1: Word classification} \label{sec:word_class}

\begin{table}[t]
     \caption{Word classification accuracy (\%) using nearest neighbor search between audio embeddings and text embeddings for varying search vocabulary size and embedding dimensions}
     \label{tab:word_recog}
     \centering
     \begin{tabular}{cccccc}
     \toprule
     Method  & Dimensions & 20k  & 100k & 500k & 900k \\
     \midrule
     FST-ASR & N/A        & 69.5 & 61.6 & 52.8 & 49.8 \\
     \midrule
     \multirow{8}{*}{\shortstack{Phone\\encoder}}
             & 8          & 53.9 & 40.5 & 28.4 & 24.1 \\
             & 16         & 70.1 & 58.2 & 45.8 & 41.3 \\
             & 32         & 74.8 & 64.0 & 52.5 & 47.9 \\
             & 48         & 75.5 & 64.9 & 53.0 & 48.1 \\
             & 64         & 75.4 & 64.5 & 52.8 & 48.4 \\
             & 128        & 75.5 & 64.9 & 53.0 & 48.7 \\
             & 256        & 75.3 & 64.6 & 52.5 & 48.3 \\
             & 512        & 75.3 & 64.4 & 52.2 & 48.0 \\
     \midrule
     \multirow{8}{*}{\shortstack{Grapheme\\encoder}}
             & 8          & 55.1 & 41.1 & 25.4 & 20.0 \\
             & 16         & 70.7 & 58.5 & 43.2 & 35.9 \\
             & 32         & 75.2 & 64.0 & 47.8 & 40.4 \\
             & 48         & 76.3 & 64.9 & 48.2 & 40.7 \\
             & 64         & 76.1 & 64.8 & 48.0 & 40.7 \\
             & 128        & 76.4 & 64.5 & 48.3 & 40.7 \\
             & 256        & 76.3 & 64.3 & 48.2 & 41.2 \\
             & 512        & 76.1 & 64.8 & 48.1 & 40.7 \\
     \bottomrule
     \end{tabular}
\end{table}

In an isolated word classification task, we classify a whole audio segment into 1 of $N$ words, assuming that only 1 word is spoken in the audio.
If the text embedding for each word $w_j$ is $\g_j \ (j \in [N])$ (we use the notation $[n]=\{1, 2, \cdots, n\}$ throughout this paper) and the input audio is mapped to an embedding $\f$, the Bayes decision rule applied to \eqref{eq:fg_dist} under the assumption of equal priors reduces to:
\begin{equation}\label{eq:euclidean_match}
i = \arg \max_{j \in [N]} p(\f|w_j) = \arg \min_{j \in [N]} \|\f - \g_j \|^2.
\end{equation}

For this experiment, we first compiled a random set of 10k unique ``test words'' by looking for word audio segments in the \emph{EXP} data that were at least 500ms in length.
To use as a search vocabulary, we compiled lists of random words of varying size ($N$ in \eqref{eq:euclidean_match}) -- 20k, 100k, 500k, and 900k words, which corresponded to 36k, 210k, 1M, and 1.7M unique pronunciations -- that included the test words.
With the phone encoder, we encoded the pronunciations, and with the grapheme encoder, we encoded the orthographies.
When evaluating the phone encoder, if the best-matching pronunciation was one of the pronunciations for the ground-truth word, the result was deemed correct.
With the grapheme encoder, if the best-matching orthography was an exact match with the ground-truth word, the result was deemed correct.
For comparison, we ran the same word classification experiment with the same DNN-HMM monophone hybrid model that was used as the frontend for the audio embedders, using single-word FST grammars built from the search vocabularies, and computing accuracy in the same manner as the phone encoder.
As shown in Tab. \ref{tab:word_recog}, phone embeddings with 48 dimensions matched the accuracy of FST decoding for a 500k vocabulary, and embeddings with 128 dimensions had accuracy that was 1.1\% points (2.2\% relative) less than FST decoding for a 900k vocabulary.
Such accuracy with low dimensions is especially useful for resource-constrained devices, and is also a key motivation for applying whole word embeddings to a continuous ASR system with a large interchangeable vocabulary \cite{yen-2023}.
The grapheme embeddings performed worse than phoneme embeddings for larger vocabulary sizes, primarily because they were more exposed to acoustic uncertainty in the embeddings for words that have multiple pronunciations (e.g. ``live,'' which could be pronounced as either {\fontfamily{cmr}\selectfont[l ih1 v]} or {\fontfamily{cmr}\selectfont[l ay1 v]}).
For smaller vocabulary sizes, one possible reason that the embeddings performed better than FSTs is that the embeddings were better trained for classifying speech segments taken out of context, but the effect seemed to diminish as the amount of confusion between pronunciations increased with larger vocabulary size.
The performance of the embeddings also seemed to saturate at around 128 dimensions and beyond.

\subsection{Experiment 2: OOV word recovery} \label{sec:phonetic_match}

\begin{table}[t]
     \caption{Percentage of OOV errors corrected by phonetic similarity match between the ASR's (erroneous) output and a 10k OOV word set using a phone embedder of varying dimensions}
     \label{tab:oov-recovery}
     \centering
     \begin{tabular}{lcc}
     \toprule
     Dissimilarity function       & Embedding     & Recovery \\
     between pronunciation pairs  & dimensions    & rate(\%) \\
     \toprule
     Minimum edit distance        & -             & 56.0     \\
     \midrule
     \multirow{6}{*}{\shortstack[l]{Distance between\\embeddings}} 
     & 16   & 45.9  \\
     & 32   & 54.5  \\
     & 48   & 55.4  \\
     & 64   & 55.5  \\
     & 128  & 55.4  \\
     & 256  & 55.0  \\
     \bottomrule
     \end{tabular}
\end{table}

The phonetic similarity measure in \eqref{eq:gg_dist} can be applied in an OOV word recovery \cite{serrino-2019} experiment where the output of an ASR is corrected by remapping it to the most phonetically-similar OOV word \cite{zhou22h_interspeech}.
In this simplified scenario, we built an FST-based isolated-word ASR with a ``deficient'' 940k-word vocabulary (around 1.8M unique pronunciations) that excluded the 10k test words and their pronunciations in Section \ref{sec:word_class}.
Hence, when presented with a test audio sample, the FST could never output the correct pronunciation, but ostensibly something acoustically similar (e.g. for the test word ``events {\fontfamily{cmr}\selectfont[ih0 v eh1 n t s]}'' it may output ``invents {\fontfamily{cmr}\selectfont[ih0 n v eh1 n t s]}'').
We then mapped the ASR's output pronunciation to the most similar pronunciation among the 10k test words using a nearest-neighbor search between their phone embeddings.
Accuracy was computed by counting the number of times the correct pronunciation was chosen.

As an alternative to embedding distances, we also tried using the well-known minimum edit distance between phone sequences to represent phonological dissimilarity \cite{kessler-1995} \cite{ghannay-etal-2016-evaluation,abdullah-2021}.
One disadvantage of the edit distance is that it assumes a static context-independent substitution cost \cite{audhkhasi-2007} between every pair of phones, hence ignoring coarticulation effects that can cause phones to sound different based on their surrounding phones \cite{kent-1977}.
The insertion and deletion costs are also set arbitrarily.
In contrast, a neural network encoder can see the entire word before outputting an acoustic embedding, which enables it to capture the internal context-dependencies of the word in a holistic representation, not to mention that computing a distance between two vectors is orders of magnitude faster than the edit distance between two sequences.
On the other hand, like all statistical models, the encoders are subject to modeling errors arising from defects in transcription, G2P, and segmentation, and model mismatch between training and test data.
An embedder trained using a word length distribution shown in Fig. \ref{fig:pronlendist} may not perform well with long strings like mailing addresses.

To obtain the substitution costs for the minimum edit distance, we constructed a phone confusion matrix by running single phones through the phone encoder and calculating their pairwise distances.
The cost of substituting phone $w_i$ with phone $w_j$ (and vice versa) is defined as:
\begin{equation}\label{eq:subst_cost}
1-2s(w_i,w_j; \lambda),
\end{equation}
where $s(w_i, w_j; \lambda)$ is defined in \eqref{eq:gg_dist} and we use 64-dimensional phone embeddings \footnote{It turns out, however, that using a binary (0 or 1) substitution cost gives identical results because $s(w_i, w_j; \lambda) \ll 0$}.
Identical phone pairs would have cost 0 while maximally different phone pairs would have cost 1.
All deletion and insertion costs were set to 1.
Tab. \ref{tab:oov-recovery} shows the percentage of words output by the ASR that were successfully mapped to the correct OOV word.
The recovery rate increased as we increased the embedding dimensions, reaching 55.5\% with 64 dimensions compared to an edit-distance-based recovery rate of 56.0\%.

\subsection{Experiment 3: Dialect clustering} \label{sec:dialect_analysis}

\begin{figure}
     \centering
     \includegraphics[width=0.35\textwidth]{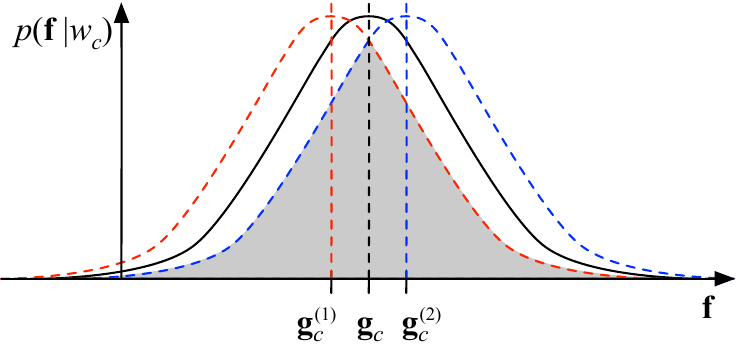}
     \caption{Hypothesized distribution shift for two different dialects of English for a given word $w_c$. If the embedding centroid for $w_c$ over general speakers is $\g_c$, we hypothesize that each dialect manifests as a shift of the general distribution in different directions, resulting in centroids $\g_c^{(1)}$ and $\g_c^{(2)}$ for Dialects 1 and 2, respectively. The (shaded) area of overlap between the two shifted distributions is the overall phonetic similarity between the two dialects' examples of word $w_c$.}
     \label{fig:dialect-shift}
\end{figure}

\begin{table}[t]
     \caption{Pairwise dialect dissimilarity computed by embedding (64 dimensions) distances for TIMIT \texttt{sa1} utterances for 6 dialects according to Eq. \eqref{eq:dissimilarity}. The full matrix is symmetric with zero diagonals, so only the lower-triangular part is shown. The dialect regions dr1, dr2, dr3, dr4, dr5, dr7 are New England, Northern, North Midland, South Midland, Southern, and Western, respectively.}
     \label{tab:dissimilarity}
     \centering
     \begin{tabular}{cccccc}
     \toprule
         & dr1     & dr2     & dr3     & dr4     & dr5    \\
     \midrule
     dr2 & 0.61572 & *       & *       & *       & *       \\
     dr3 & 0.58936 & 0.13495 & *       & *       & *       \\
     dr4 & 0.65666 & 0.33050 & 0.26793 & *       & *       \\
     dr5 & 0.60411 & 0.37564 & 0.34027 & 0.18538 & *       \\
     dr7 & 0.63036 & 0.14484 & 0.10431 & 0.30236 & 0.36269 \\
     \bottomrule
     \end{tabular}
\end{table}

\begin{figure}[t]
     \centering
     \begin{subfigure}{0.5\textwidth}
          \centering
          \includegraphics[width=0.85\textwidth]{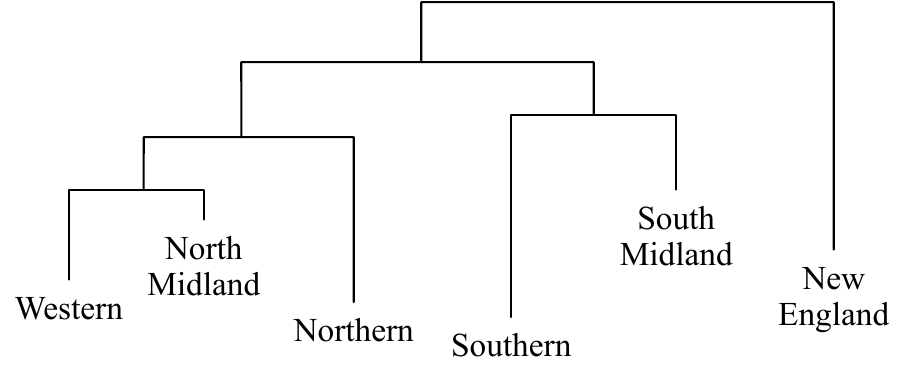}
          \setlength{\abovecaptionskip}{0in}
          \caption{$d=64$}
          \label{fig:addtree-identical}
     \end{subfigure}%
     \hfill
     \vspace{.05in}
     \begin{subfigure}{0.5\textwidth}
          \centering
          \includegraphics[width=0.85\textwidth]{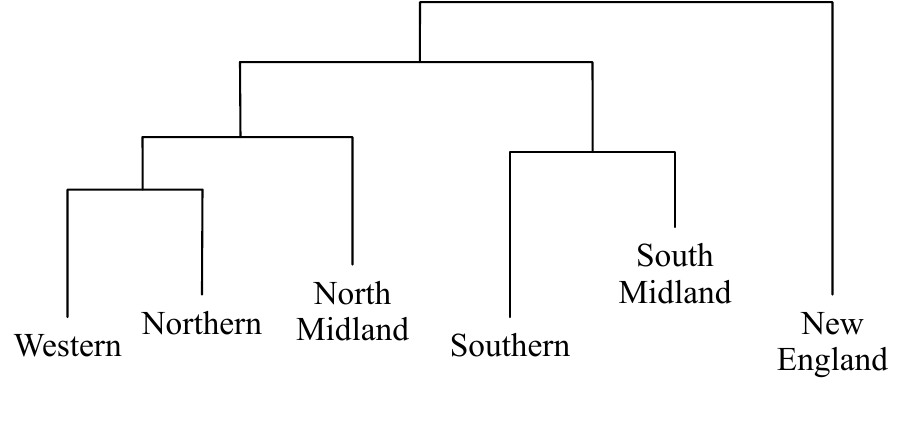}
          \setlength{\abovecaptionskip}{-.15in}
          \caption{$d=48$}
          \label{fig:addtree-similar}
     \end{subfigure}
     \caption{ADDTREE/P \cite{addtree} clustering result (not accurately drawn to scale) of the pairwise dialect similarity matrices computed using audio embeddings. With 64-dimensional embeddings, the cluster hierarchy is identical to that obtained from human listening experiments in Clopper et al. \cite{CLOPPER2004111} (Fig. 4 therein). For $d=48$, the hierarchy is near-identical with only Northern and North Midland being swapped.} 
     \label{fig:addtree}
\end{figure}

Another way to see whether our embeddings are arranged in a meaningful way is to investigate whether specific clusters of the embeddings form a relationship that is consistent with other findings in the literature.
Given two groups of speakers speaking two different dialects, all uttering the same word $w_c$, suppose we obtain $n$ audio embedding samples $F_1=\{ \f_1^{(1)}, \cdots, \f_n^{(1)} \}$ from the first population and $m$ audio embedding samples $F_2=\{ \f_1^{(2)}, \cdots, \f_m^{(2)} \}$ from the second.
As shown in Fig. \ref{fig:dialect-shift}, we hypothesize that the distribution of embeddings for each dialect manifests as a shift of the general population's distribution.
The sample means over $F_1$ and $F_2$ are the dialect-wise centroids per Eq. \eqref{eq:mean}, $\g_c^{(1)}$ and $\g_c^{(2)}$.
The area of overlap per Eq. \eqref{eq:gg_dist} using $\g_c^{(1)}$ and $\g_c^{(2)}$ would represent the phonetic similarity between the two dialects for word $w_c$.
When there are more than 2 dialects, we can compute the pairwise similarity between all possible pairs of dialects and apply any known clustering or data visualization techniques to see how the dialects differ from each other.

Of particular interest to us is the study by Clopper et al. \cite{CLOPPER2004111} which measured the perceptual dissimilarity between different dialects using human listeners.
Labeled data from six dialects in the United States in the TIMIT\cite{garofolo-1992} corpus was used in the study.
In particular, \cite{CLOPPER2004111} showed clustering hierarchies of the dialects that were consistent across all listener groups when using audio data for the \texttt{sa1} sentence, so we used the same sentence in our experiment.
For each of the 11 words in ``{\fontfamily{cmr}\selectfont SHE HAD YOUR DARK SUIT IN GREASY WASH WATER ALL YEAR}'' we computed six dialect-wise audio embedding centroids.
The dissimilarity $d(i,j)$ between dialect $i$ and dialect $j$ ($i, j \in [6]$) is the mean dissimilarity between the centroids across the 11 words where, as in Eq. \eqref{eq:subst_cost}, we turn the similarity in Eq. \eqref{eq:gg_dist} into a dissimilarity:
\begin{equation}\label{eq:dissimilarity}
     d(i,j) = \frac{1}{11} \sum_{c=1}^{11} \left[ 
     1- \exp \left\{ -\frac{1}{8\sigma^2} \Big\| \g_c^{(i)} - \g_c^{(j)} \Big\|^2 \right\}
     \right]
\end{equation}
where $\g_c^{(i)}$ is the centroid for the $c$'th word $(c \in [11])$ in the $i$'th dialect $(i \in [6])$.
Table \ref{tab:dissimilarity} shows the lower-triangular part of the dialect dissimilarity matrix (the full matrix is symmetric with diagonals being 0) when using the 64-dimensional audio embedder.
Same as in \cite{CLOPPER2004111}, we ran the matrix through ADDTREE/P\cite{addtree}\cite{addtree_pas} to obtain a visualization of clusters in Fig. \ref{fig:addtree}.
We can see the tree hierarchy (ignoring specific branch lengths) for $d=64$ is identical to those in \cite{CLOPPER2004111} (Fig. 4 therein) that were obtained via human listening experiments, and near-identical with only minor differences in the case of $d=48$.
Although not shown here, the hierarchy was also near-identical for $d=32$ and $d=128$. 
If our embeddings-based results match those of human listener-based experiments such as \cite{CLOPPER2004111}, this would serve as another form of validation of our theoretical framework as well as inspire further cross-domain application of acoustic word embeddings to linguistics, audiology, and other related fields.

\subsection{Experiment 4: Wake-up word confusion} \label{sec:wakeup}

\begin{figure}[t]
     \centering
     \begin{subfigure}{0.49\textwidth}
          \centering
          \includegraphics[width=0.97\textwidth]{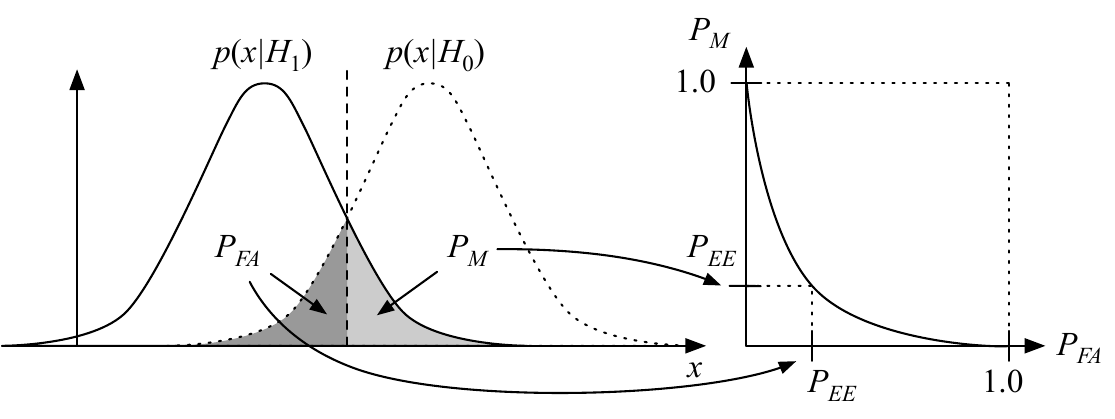}
          \caption{Fig. \ref{fig:overlap} interpreted as a detection problem where the hypothesis is word $w_1$ and the null hypothesis is word $w_2$. Assuming equal priors and equal-variance Gaussian distributions, the overlapping area (phonetic similarity) directly becomes the equal error rate ($P_{EE}$) in the receiver operating characteristic (ROC) curve.}
          \label{fig:detection}
     \end{subfigure}%
     \hfill
     \vspace{-0.1in}
     \begin{subfigure}{0.49\textwidth}
          \centering
          \includegraphics[width=0.97\textwidth]{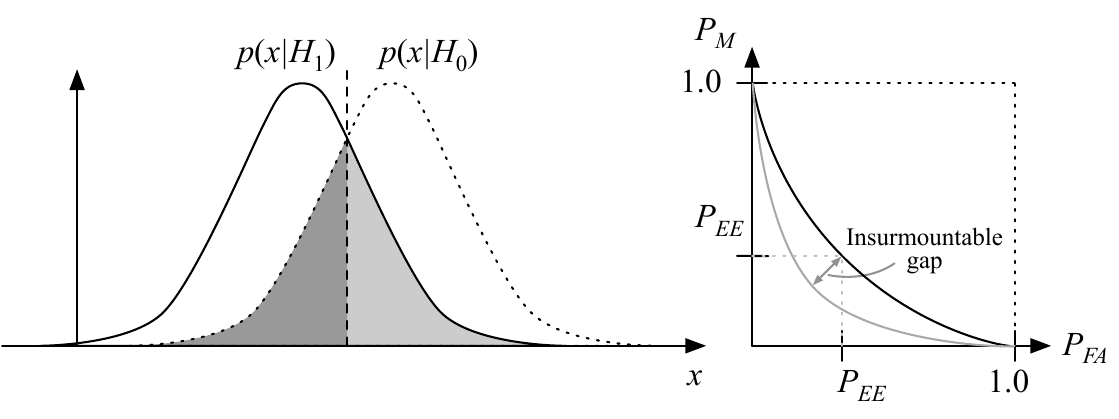}
          \caption{When there is more overlap between words $w_1$ and $w_2$, the equal error rate increases and we get a worse ROC curve than the one in Fig. \ref{fig:detection}. Because the Bayes Error Rate is a theoretical bound, the gap between the two ROC curves is insurmountable.}
          \label{fig:largeoverlap}
     \end{subfigure}
     \caption{Interpretation of phonetic similarity in a detection setting.}
\end{figure}

\begin{figure}[t]
     \centering
     \begin{subfigure}{0.5\textwidth}
          \centering
          \includegraphics[width=0.85\textwidth]{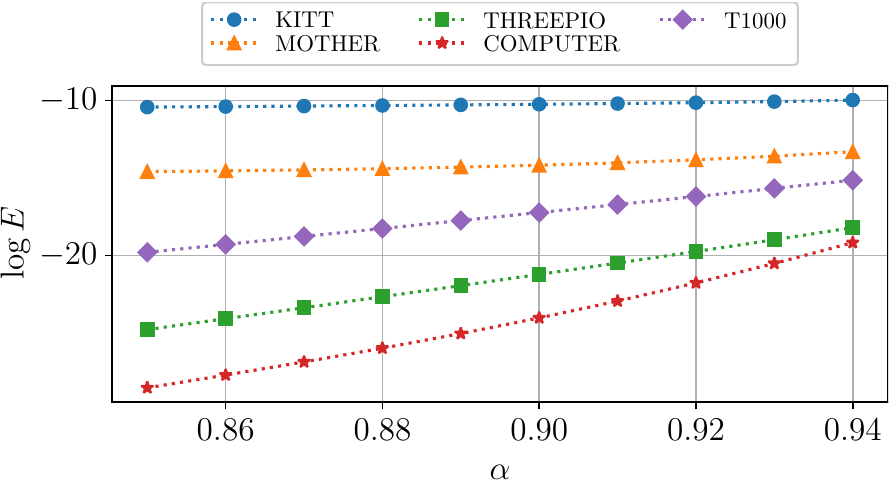}
          \caption{Grapheme embedder with 64 dimensions}
          \label{fig:wakeup1}
     \end{subfigure}%
     \hfill
     \vspace{0.1in}
     \begin{subfigure}{0.5\textwidth}
          \centering
          \includegraphics[width=0.85\textwidth]{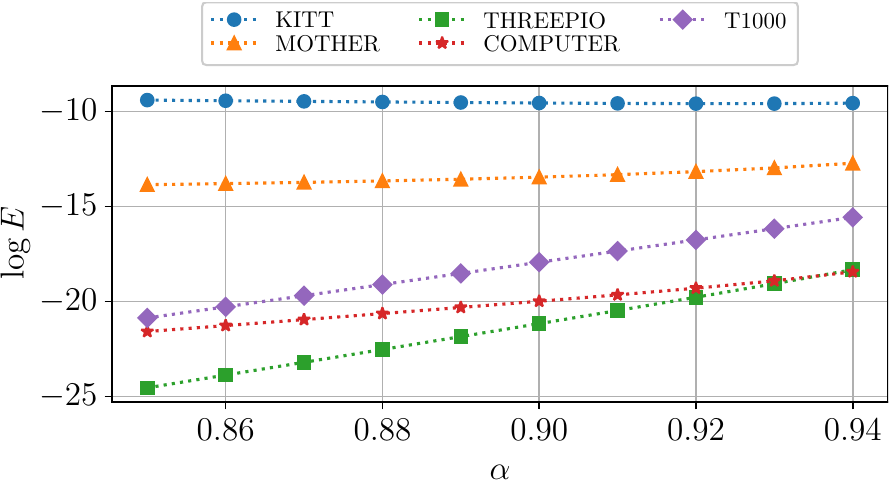}
          \caption{Alternate grapheme embedder}
          \label{fig:wakeup2}
     \end{subfigure}
     \caption{Log expected confusion per Eq. \eqref{eq:logsumexp} for different wake-up words for varying LM scale using two different 64-dimensional grapheme embedders and a trigram LM.}
     \label{fig:wakeup}
\end{figure}

Modern consumer devices with speech-based interfaces often support a ``wake-up word'' that a user can utter to activate the device\cite{heysiri}.
However, wake-up word detection is prone to accidental triggers \cite{SCHONHERR2022101328}.
It is natural to postulate that certain wake-up words would be inherently more likely to cause accidental triggers than others.
``Anne,'' for example, would more easily result in accidental triggers than ``Anastasia,'' since the former can be confused with the oft-spoken ``and,'' and would be a poor choice for a wake-up word.
Tuning the system to trigger more conservatively would in turn cause an increase in false rejections, so an overall degradation in accuracy would always be expected.
The theoretical framework proposed in this paper allows us to quantitatively predict the relative accuracy for different wake-up words, which may be useful when choosing a wake-up word for a new device.

First, it is well-known that a 2-class classification problem can be mapped to a detection problem.
Returning to Fig. \ref{fig:overlap}, we consider the hypothesis $H_1$ as the presence of word $w_1$ and the null hypothesis $H_0$ as the presence of word $w_2$.
Using $P(w|x)p(x)=p(x|w)P(w)$ and assuming equal priors $P(w_1)=P(w_2)$, the overlapping area in Fig. \ref{fig:overlap} directly maps to the equal error rate $P_{EE}$ in the receiver operating characteristic (ROC) in the detection problem, shown in Fig. \ref{fig:detection}.
If we chose a different $w_1$ and $w_2$ that have greater phonetic similarity, we would have higher $P_{EE}$ and a worse ROC curve as in Fig. \ref{fig:largeoverlap}.
Since the Bayes Error Rate is a theoretical bound, it should also be apparent that the performance gap in the two ROC curves is insurmountable; the ideal detector for the words in Fig. \ref{fig:largeoverlap} will never be as good as the ideal detector for the words in Fig. \ref{fig:detection}.

Using pairwise similarities $s(\cdot,\cdot;\lambda)$ defined in Eq. \eqref{eq:gg_dist} according to our embedding model $\lambda$, we can compute an expected pairwise confusion over all words in the vocabulary as an indication of how likely an accidental trigger can occur for a given wake-up word $w_t$.
Note that we are not considering accidental triggers due to non-speech sounds:
\begin{equation}
     E_{W,W \ne w_t}[s(w,w_t;\lambda)]=\sum_{w \ne w_t} s(w_t,w) P(w).
\end{equation}
Above, $P(w)$ is a LM score. Taking the log and applying the time-honored LM scale $\alpha \in [0,1]$ to submit to the reality that models never give true probabilities, we have:
\begin{multline}\label{eq:logsumexp}
     \log E_{W,W \ne w_t}[s(w,w_t;\lambda)] = \\
     \underset{w, w \ne w_t}{\mathrm{LSE}}
      \left[ 
      (\alpha-1) \left(
          \frac{\| \g_t - \g \|^2}{8\sigma^2} + \log 2
          \right)
     + \alpha \log P(w)
      \right],
\end{multline}
where $\mathrm{LSE}$ stands for $\mathrm{LogSumExp}$.
An automated way of tuning $\alpha$ is beyond the scope of this paper.
For this experiment, we empirically found the range 0.85 to 0.95 to give a reasonable balance between phonetic similarity and linguistic likelihood of occurrence by inspecting examples of nearest neighbors resulting from the interpolated similarity function inside the $\mathrm{LSE}$ operator above.
Fig. \ref{fig:wakeup1} shows the log expected confusion for 5 hypothetical wake-up words\footnote{These can be easily changed in the source code} over a range of values of $\alpha$ surrounding 0.9.
An overall trend can be seen where ``Kitt'' has the most confusion followed by ``Mother'', ``Tee One Thousand'', ``Threepio'', and ``Computer.''

Some caution must be taken when we view these results.
First, how an actual wake-up word detection model performs depends greatly on model structure and training methodology and data.
If both a prediction model and the actual model are reasonably developed, however, we can expect a correlation to exist between their results.
Second, the expected confusion is itself based on a model that estimates but can never truly reflect reality.
Hence, it is important to apply this paper's proposed theoretical framework to diverse applications so that we may gain greater confidence that its predictions are useful.
Also, predictions from multiple embedders and LMs should be inspected, and the domain and characteristics of their training data should be taken into account.
To further demonstrate this, we reran the experiment using an alternate grapheme embedder trained using a different, less-scrubbed subset of LibriHeavy, shown in Fig. \ref{fig:wakeup2}, and we can see that the rank changed for two of the wake-up words for lower values of $\alpha$.

In future applications, the metric in \eqref{eq:logsumexp} could be used to find specific words in the vocabulary that have high confusion with a wake-up word, which could be used as adversarial samples \cite{wang22c_odyssey} for fine-tuning wake-up models.
One could also exhaustively mine a large database of transcribed audio to look for speech segments with audio embeddings that are close to the text embedding of the wake-up word but have a different transcription and use them as adversarial audio samples.

\section{Conclusion}

We have proposed a theoretical framework for interpreting acoustic neighbor embeddings.
We showed how the distance between an audio embedding and a text embedding reflects a Gaussian likelihood score, and the distance between text embeddings reflect the area of overlap between two distributions that define phonetic similarity.
Theoretical and empirical evidence was shown for making an approximation of uniform cluster-wise isotropy, which enables us to use Euclidean distances in all our applications.
Four different experiments that demonstrate and partially validate the framework were discussed.

\section{Future Work}

The LSTMs used for all our embedders could be replaced by attention-based models \cite{vaswani2023attentionneed} for improved performance, particularly for word length distributions with longer tails than in Fig. \ref{fig:pronlendist}.
No attempt has been made in this work to compare the proposed embeddings with others in the literature \cite{bengio14_interspeech,levin_2013,settle17_interspeech,7178970,7472619,7846310,palaskar-2019,8736337,9360516,he-2017}.
It is unknown whether the proposed theoretical framework would be valid for other embeddings due to fundamental differences in how they are trained (e.g. using a contrastive loss \cite{7472619}).
If we wish to answer that question, further study and experimentation would be needed.

\section{Acknowledgements}

The author would like to thank Erik McDermott, Zak Aldeneh, and Barry Theobald for helpful comments that greatly improved the quality of this manuscript.

\appendices{
\section{Further details on embedder training}\label{sec:app-training}
Assume that we have 10 total training samples $(A_i, B_i, C_i), \ i \in [10]$ as follows:

\begin{table}[H]
     \centering
     \begin{tabular}{clll}
     \toprule
     $i$ & $A_i$ & $B_i$ & $C_i$ \\ 
     \midrule
     1  & $A_1$  & [ah0 b eh1 t]   & abet   \\
     2  & $A_2$  & [k r ae1 b]     & crab   \\
     3  & $A_3$  & [k r ae1 b]     & crab   \\
     4  & $A_4$  & [k r ae1 b]     & crab   \\
     5  & $A_5$  & [f ae1 k t]     & fact   \\
     6  & $A_6$  & [s ae0 k]       & sack   \\
     7  & $A_7$  & [s ae1 k]       & sack   \\
     8  & $A_8$  & [w ey1 t]       & wait   \\
     9  & $A_9$  & [w ey1 t]       & weight \\
     10 & $A_{10}$ & [z iy1 b r ah0] & zebra  \\
     \bottomrule
     \end{tabular}
\end{table}
An example microbatch of size $M=4$ would be $\{A_2, A_3, A_5, A_8\}$ where the pivot is $A_2$, a ``same'' sample is $A_3$, and the rest are ``different'' samples $A_5$ and $A_8$. We have $p_{01}=1$ and $p_{0,2}=p_{0,3}=0$ in the simplified loss in Eq. \eqref{eq:kl_dist_simple}. More examples are as follows:
\begin{table}[H]
     \centering
     \begin{tabular}{clllll}
     \toprule
     Pivot & Same & Different & $p_{01}$ & $p_{02}$ & $p_{03}$ \\ 
     \midrule
     $A_2$ & $A_3$        & $A_5$, $A_8$  & 1.0 & 0   & 0 \\
     $A_8$ & $A_9$        & $A_2$, $A_3$  & 1.0 & 0   & 0 \\
     $A_3$ & $A_2$, $A_4$ & $A_7$         & 0.5 & 0.5 & 0 \\
     \bottomrule
     \end{tabular}
\end{table}
It should be evident that the audio samples $A_1$, $A_5$, $A_6$, $A_7$ $A_{10}$ can never be pivots since no ``same'' samples for them exist in the data.
Alternate implementations may choose to ignore phone stress markers, regarding $A_6$ and $A_7$ as ``same.''
The grapheme sequences $C_i$'s play no role in our audio embedder training, but alternate implementations may assign ``same'' or ``different'' based on the $C_i$'s \cite{he-2017} rather than the $B_i$'s in Eq. \eqref{eq:dist}.

\section{Further details on the approximation of constant cluster-wise isotropy} \label{sec:app-isotropy}

\begin{figure}
    \centering
    \includegraphics[width=0.4\textwidth]{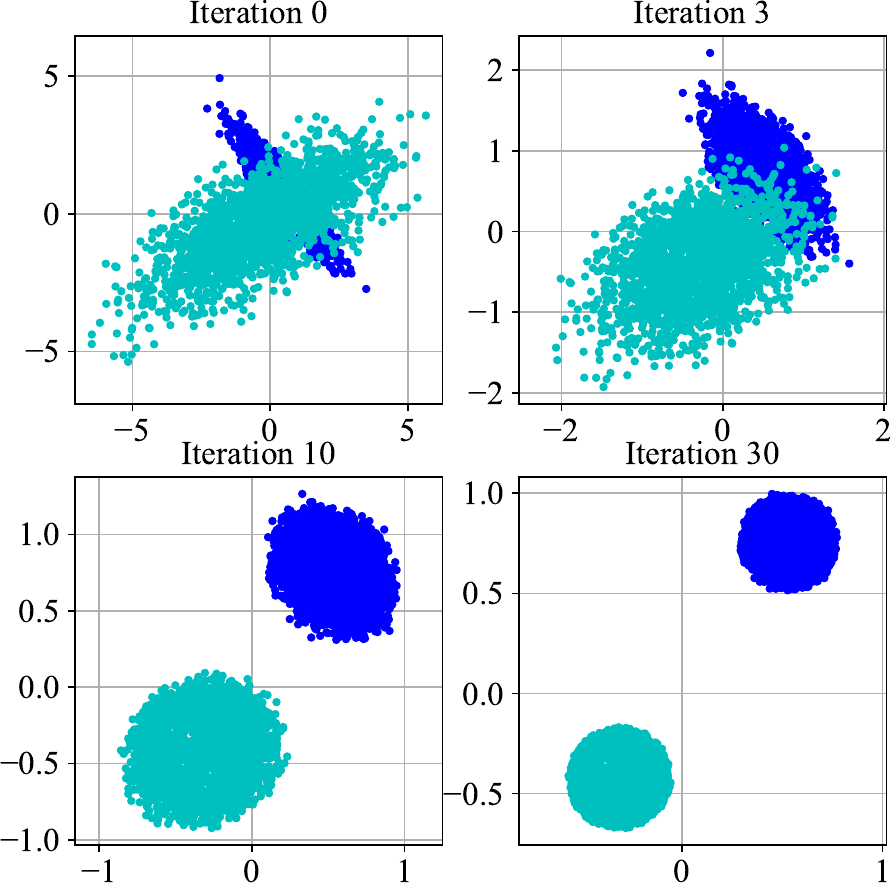}
    \caption{Simulation of embeddings for two clusters being trained according to the gradient descent equation in \eqref{eq:backprop}. The clusters are initially scattered in very different directions and widths, but with enough iterations become equal-sized spheres.}
    \label{fig:simul}
\end{figure}

We first inspect the gradient of the training loss function for the audio encoder $f(\cdot)$, which can be obtained by applying \eqref{eq:neighbor_prob2} to \eqref{eq:sne_gradient}.
The summation can be split into the sum over the ``positive'' samples with the same subword sequence as $i$, i.e., $J_i^+ = \left\{ j: B_j = B_i \right\}$, and the ``negative'' samples, i.e., $J_i^- = \left\{ j : B_j \ne B_i \right\}$:
\begin{equation}\label{eq:spe_gradient}
\frac{\partial \L_{f}}{\partial \f_i} = \frac{\partial \L_{f}^+}{\partial \f_i} + \frac{\partial \L_{f}^-}{\partial \f_i}
\end{equation}
where
\begin{align}
\frac{\partial \L_{f}^+}{\partial \f_i} &= 2 \sum_{j \in J_i^+} (\f_i - \f_j) \left( \frac{2}{c_{i}}- q_{ij} - q_{ji} \right) \label{eq:gradient_plus} \\
\frac{\partial \L_{f}^-}{\partial \f_i} &= - 2 \sum_{j \in J_i^-} (\f_i - \f_j) (q_{ij} + q_{ji}). \label{eq:gradient_minus}
\end{align}
For simplicity, we assume that we have already reached a training state where $\f_i$ is sufficiently separated from the negative samples in $J_i^-$ such that $q_{ij}$ and $q_{ji}$ are close to 0 for $j \in J_i^-$.
This allows us to assume ${\partial \L_{f}^-} / {\partial \f_i} \approx 0$.

\begin{figure}[t]
     \centering
     \begin{subfigure}{0.24\textwidth}
          \centering
          \includegraphics[width=0.99\textwidth]{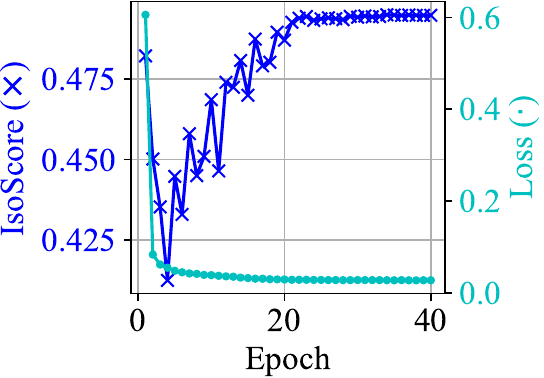}
          \caption{$d=16$}
          \label{fig:isoscore16}
     \end{subfigure}%
     \begin{subfigure}{0.24\textwidth}
          \centering
          \includegraphics[width=0.99\textwidth]{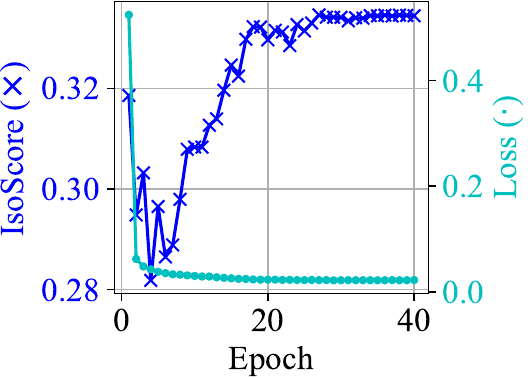}
          \caption{$d=32$}
          \label{fig:isoscore32}
     \end{subfigure}
     \begin{subfigure}{0.24\textwidth}
          \centering
          \includegraphics[width=0.99\textwidth]{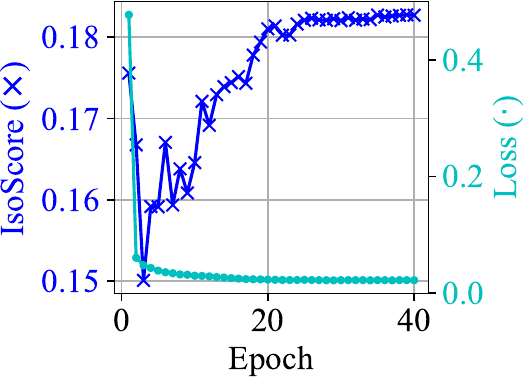}
          \caption{$d=64$}
          \label{fig:isoscore64}
     \end{subfigure}%
     \begin{subfigure}{0.24\textwidth}
          \centering
          \includegraphics[width=0.99\textwidth]{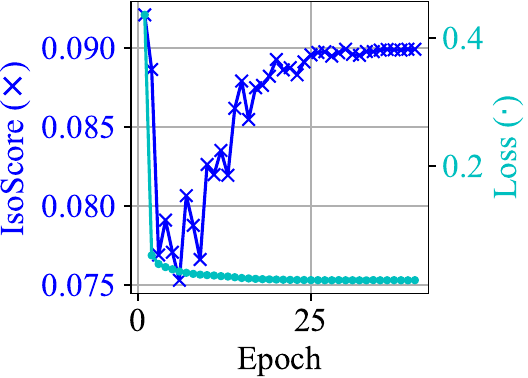}
          \caption{$d=128$}
          \label{fig:isoscore128}
     \end{subfigure}
     \caption{Average cluster-wise IsoScore \cite{rudman-etal-2022-isoscore} (marked by $\times$) for 490 clusters and the mean loss (marked by $\cdot$) over 40 epochs of actual audio embedder training for $d$-dimensional embeddings. Higher IsoScore implies greater isotropy.}
     \label{fig:isoscore}
\end{figure}
\begin{figure}[t]
     \centering
     \begin{subfigure}{0.24\textwidth}
          \centering
          \includegraphics[width=0.99\textwidth]{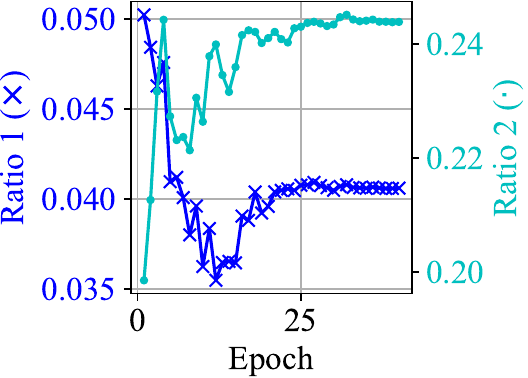}
          \caption{$d=16$}
          \label{fig:ratio16}
     \end{subfigure}%
     \begin{subfigure}{0.24\textwidth}
          \centering
          \includegraphics[width=0.99\textwidth]{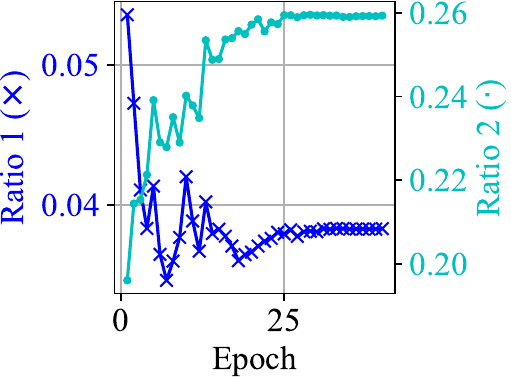}
          \caption{$d=32$}
          \label{fig:ratio32}
     \end{subfigure}
     \begin{subfigure}{0.24\textwidth}
          \centering
          \includegraphics[width=0.99\textwidth]{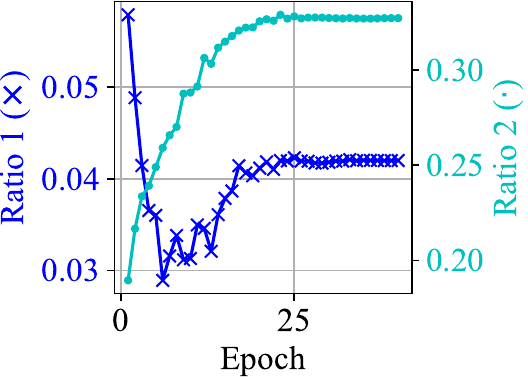}
          \caption{$d=64$}
          \label{fig:ratio64}
     \end{subfigure}%
     \begin{subfigure}{0.24\textwidth}
          \centering
          \includegraphics[width=0.99\textwidth]{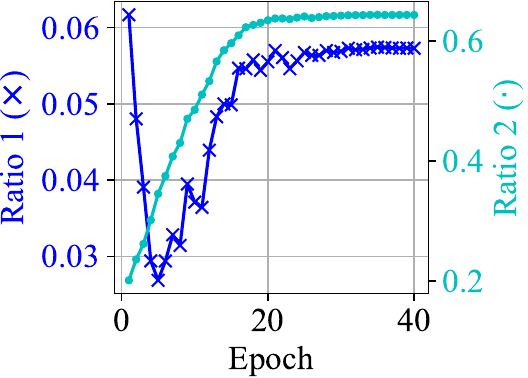}
          \caption{$d=128$}
          \label{fig:ratio128}
     \end{subfigure}
     \caption{Two ratios for measuring the degree of uniformity in intra-cluster variance, for the same clusters in Fig. \ref{fig:isoscore} for $d$-dimensional embeddings. Lower ratios imply more uniform intra-cluster variances. Compared to isotropy, it is harder to prove an increase in uniformity, especially for larger $d$.}
     \label{fig:ratio}
\end{figure}

If we assume that the neural network is capable of modeling each embedding $\f_j$ with unlimited freedom, we can assume that at every step in backpropagation training, each embedding will directly follow the gradient descent rule
\begin{align}
\f_i' & = \f_i - \gamma \frac{\partial \L_{f}}{\partial \f_i} \approx \f_i - \gamma \frac{\partial \L_{f}^+}{\partial \f_i} \label{eq:backprop}
\end{align}
where $\f_i'$ is the updated embedding and $\gamma$ is a learning rate.

Two observations can be made in \eqref{eq:backprop}: 
First, the farther $\f_i$ is from the other points in the cluster, the smaller the values of $q_{ij}$ and $q_{ji}$ and therefore the larger the change in $\f_i$ at each step.
Hence, initially-far points will eventually "catch up" to the initially-close points and become equally close to the cluster centroid. 
Second, other than the proximity of $\f_i$ to the other points in the cluster, there is no other external factor that affects the step size.
Hence, points that have similar distance to the centroid will tend to move at the same rate, irrespective of their initial locations.
This is especially due to the fact that we have assumed binary distances in \eqref{eq:dist} and \eqref{eq:neighbor_prob2}, and there is no explicit variable distance as in \eqref{eq:euclidean_dist}.

Fig. \ref{fig:simul} is a simulation where two artificially-generated clusters of embeddings are subject to \eqref{eq:backprop} with $\gamma=0.1$.
Even though the clusters are initially scattered in different directions at different lengths, by the 30'th iteration both clusters become spheres, as a consequence of the first observation above.
Furthermore, the size of the spheres eventually become equal, as a consequence of the second observation.
An even distribution of points in all directions implies isotropic covariance.
If we also allow the radius of the spheres to represent the size of cluster scatter (which, strictly speaking, is not necessarily true because the boundaries alone do not give a full picture of how the points are distributed \emph{inside} the sphere), the equal-size scatter also implies equal variance.

To observe the first effect, isotropic covariance, in our actual embeddings (obtained in Section \ref{sec:experiments}), we measure the average \emph{cluster-wise} IsoScore \cite{rudman-etal-2022-isoscore} for our $\f$ embeddings for different epochs during actual audio embedder training.
In previous studies, the IsoScore was used to measure the \emph{global} isotropy of word embeddings \cite{rudman-etal-2022-isoscore,abdullah-2022}.
As shown in Fig. \ref{fig:isoscore}, as the training progresses and the loss decreases, the average cluster-wise IsoScore increases.
The IsoScore also tends to be higher for lower-dimension embeddings.

The second effect, equal variance, is harder to find.
If we assume isotropy, the value of $\sigma$ in \eqref{eq:iso_bound} for each cluster is the average element-wise standard deviation of $\f$, computed over all instances of $\f$ for the same word.
As a measure of how much $\sigma$ varies from cluster to cluster, we can again compute the standard deviation of $\sigma$ across different clusters.
To remove any trivial scaling effects in the embeddings, we can divide the standard deviation by either the average element-wise absolute value of $\f$, or the average $\sigma$ across clusters.
The desired result would be to see both ratios become smaller as training progresses.
We see this happen to the former ratio in Fig. \ref{fig:ratio}, but not the latter.

\bibliographystyle{IEEEtran}
\bibliography{refs}

\end{document}